\documentclass[pdflatex,sn-mathphys-num,iicol]{sn-jnl}%

\usepackage{graphicx}%
\usepackage{multirow}%
\usepackage{amsmath,amssymb,amsfonts}%
\usepackage{amsthm}%
\usepackage{mathrsfs}%
\usepackage[title]{appendix}%
\usepackage{xcolor}%
\usepackage{textcomp}%
\usepackage{manyfoot}%
\usepackage{booktabs}%
\usepackage{algorithm}%
\usepackage{algorithmicx}%
\usepackage{algpseudocode}%
\usepackage{listings}%

\theoremstyle{thmstyleone}%

\theoremstyle{thmstyletwo}%

\theoremstyle{thmstylethree}%

\begin{document}

\title[The Hubble tension as an effect of the renormalization of fundamental constants and cosmological parameters]{The Hubble tension as an effect of the renormalization of fundamental constants and cosmological parameters}

\author*[1,2,3]{\fnm{Fabio} \sur{Briscese}}\email{fabio.briscese@uniroma3.it, briscese.phys@gmail.com}

\affil*[1]{\orgdiv{Department of Architecture}, 
	\orgname{Roma Tre University}, \orgaddress{\street{Largo Giovanni Battista Marzi, 10}, \city{Rome}, \postcode{00153}, 
		\country{Italy}}}

\affil[2]{
	\orgname{Istituto Nazionale di Fisica Nucleare, Sezione di Roma 3}, \orgaddress{\street{ Via della Vasca Navale 84}, \city{Rome}, \postcode{00146}, 
		\country{Italy}}}

\affil[3]{
	\orgname{Istituto Nazionale di Alta Matematica Francesco		Severi, Gruppo Nazionale di Fisica Matematica}, \orgaddress{\street{Citt\`{a} Universitaria, P.le A. Moro 5}, \city{Rome}, \postcode{00185}, 
		\country{Country}}}


\abstract{		In this Letter we propose an interpretation of the Hubble tension as an effect of the scale-running of cosmological parameters and fundamental constants, as the Newton's constant  $G$ and the cosmological constant $\Lambda$. Namely, the tension between $H_0$ measurements by CMB and Supernovae  observations would be a consequence  of the fact that the value of  $H_0$ depends (due to the running of $G$ and $\Lambda$) on the scale at which it is measured. Indeed, the two different values of the Hubble parameter \textit{today} correspond to the two  different cosmological scales at which Supernovae and CMB measurements probe our universe. We discuss some possible theoretical scenarios in which the running of  $G$ and $\Lambda$  can occur. We stress that one should interpret such a running as a scale-dependence of the cosmological parameters and fundamental constants at the same time rather than a time dependence, as it is usually considered in the literature.
}

\maketitle
The standard $\Lambda$-CDM model, based on a nearly spatially flat, homogeneous and isotropic  universe containing the fields of the standard model of particles, plus a large amount of dark matter and dark energy, has provided a successful description of the universe at large cosmological scales, see e.g. \cite{Mukhanov} for an introductory treatment. However, in recent years high precision cosmology has challenged the validity of this concordance model, as tensions in the measurements of cosmological parameters by means of different cosmological probes has emerged. The most cogent is the so called "Hubble tension", consisting in a difference in the measurement of the Hubble parameter today $H_0$ by CMB and supernovae observations, which has given $H_0^{cmb} = \left(67.4\pm0.5 \right) km \, s^{-1}\, Mpc^{-1}$  and \mbox{$H_0^{sn} = \left(73.17\pm 0.86\right) km \, s^{-1}\, Mpc^{-1}$} respectively, indicating a discrepancy ath a $5-6 \, \sigma$ level; while a milder, but significant, tension in  the $S_8$ parameter is also observed at a $2-3 \, \sigma$ level, see \cite{di valentino} for a review of cosmological tensions and a list of candidate solutions.

In this paper we propose a new idea:   the Hubble  tension would be due to the \textit{scale-running} of cosmological parameters and fundamental constants. In fact, it is reasonable to expect, both from fundamental physics and from coarse graining of small-scale inhomogeneities, that fundamental constants and cosmological parameters, e.g. the Newton's constant $G$ and the cosmological constant $\Lambda$, receive scale-dependent corrections, so that they have to be considered as scale-running quantities. Indeed, measurements of $G$ and $\Lambda$ at different cosmological scales would yield different values, also affecting the measured values of $H_0$. Thus, the difference between  $H_0^{cmb}$ and $H_0^{sn}$ would be  due to the fact that CMB and supernovae observations probe the universe  at two different cosmological scales. 

It is necessary to clarify that the idea of varying fundamental constants is not new, see again \cite{di valentino} and references therein. However, this has always been considered a as a time-variation, while we are proposing  that the value of cosmological parameters and fundamental constants differ  when measured \textit{at the same time} but at different cosmological scales. In fact, even when the running of cosmological parameters has been derived by means of renormalization of matter fields, e.g. as an energy-scale dependent vacuum energy as in \cite{casimir1,casimir2,casimir3,pera1,pera2,pera3,pera 4,pera 5,pera 6,pera 7,pera 8,pera H 1,pera H 2}
, such  quantities have been treated as  time dependent parameters in the Friedmann equations (through the Hubble parameter $H(t)$) for the background evolution of the universe; which is a quite different interpretation of the scale-running than the one proposed in this letter.
More precisely, the calculation of the quantum effects in \cite{casimir1,casimir2,casimir3,pera1,pera2,pera3,pera 4,pera 5,pera 6,pera 7,pera 8,pera H 1,pera H 2}
 is first performed at an arbitrary renormalization scale $\mu$, and then  this scale is associated  with the value of $H$ at each point of the cosmological expansion.

In what follows we argue how a scale-running of fundamental constants and cosmological parameters can solve the Hubble tension, and we describe different scenarios in which  such a running occurs. 
Before proceeding, we briefly discuss the paradigmatic example of the running of the electron charge in QED, \cite{peskin}. In this theory, loop contributions to scattering amplitudes contain divergences that have to be regularized first, and then renormalized, so that  the bare electron mass, charge and field strengths are replaced by their physical values at a given scale. As a result, physically measurable quantities show a dependence on the renormalization scale that encodes the effects of the short-scale physics. For instance, the renormalized electron charge has the well known form \cite{peskin}

\begin{equation}
	e^2(\mu)= \frac{e^2(\mu_r)}{1-\frac{e^2(\mu_r)}{6 \pi^2} \ln\left(\frac{\mu}{\mu_r}\right)} \, ,
\end{equation}
where $\mu_r$ is the renormalization energy-scale and $\mu$ is the energy at which the charge is measured. This implies that the value of the electron charge depends on the energy at which it is measured, entailing the growth of $e(\mu)$ at short distances, due to the large-scale screening of the bare electron charge by virtual $e^+ \, e^-$ pairs at short scales.


The renormalization group is a powerful tool, which has been developed in order to account for the unavoidable occurrence of infinities in quantum field theories \cite{peskin}, in statistical field theory \cite{parisi} and in coarse graining of systems containing more than one significant scale \cite{vulpiani,Goldberger,Burgess,Pich}. All these applications of the renormalization group share some common features that can be briefly resumed (with no aim of being exhaustive) as follows: the renormalization group provides a description of self-similar systems in terms of few parameters that depend on the  scale at which the system is considered. Such parameters account for contributions coming from smaller scales which have been integrated out. In fact, it typically happens that the system under consideration is characterized by different (two, several or infinitely many) scales, and the evolution at such scales is not decoupled. Indeed, the physics at short distances affects the properties of the system at larger scales.  Renormalization group methods allow for a description of such systems at  large scales in situations in which the short-distance contributions are divergent, and must be renormalized. Moreover, they have been proved to be capable of predicting the relation between measurable physical quantities at different scales.

The need of an effective field theory treatment of cosmological perturbations, and the consequent use of renormalization group methods for addressing unphysical divergences, has been exhaustively discussed in a series of papers \cite{EFT1,EFT2,EFT3,EFT4,EFT5}. In fact, the time dependence of the matter density contrast  implies that $\delta\rho/\rho$ can not be used as a scale-independent small parameter for perturbative series. Moreover, one has to account for large-scale deviations of the matter energy-density tensor from that of  a  perfect fluid, and handle UV divergences that occur in one-loop corrections to the density power spectrum.  In the filed treatment of cosmological perturbations, one expands in metric and velocity perturbations, which remain small even when the density contrast becomes large, and integrates out the contributions of short-scales to the dynamics of the large-scale universe.
The final result is that  short-scale inhomogeneities  introduce new dissipative terms in the dynamics  of large-scale matter perturbations, which behave as those of  a fluid  characterized by an equation of state, a sound speed and a viscosity parameter. These terms are then used as counter-terms  to cancel UV divergences in the power spectrum.

The effect of small-scale inhomogeneities on the gravitational background has  been considered  negligibly small in \cite{EFT1}  in the case of an Einstein-de Sitter universe, even if it might be at the reach of current observational precision. In a dark matter/dark energy dominated universe, one would expect that  both $G$ and $\Lambda$ would receive scale-dependent corrections from short-scale degrees of freedom. Indeed, 
we can conjecture that $G$ is a scale-running quantity that depends on the length-scale $\ell$ at which it is measured, e.g.

\begin{equation}\label{running G}
	G\left(\ell\right) =\frac{ G_r}{\left[1 + a_1 \left(\frac{\ell}{\ell_r}\right) + a_2 \left(\ln\left(\frac{\ell}{\ell_r}\right)\right)^2 +\ldots \right]} \, ,
\end{equation}
where $G_r$ is the value of the Newton's constant at the renormalization scale $\ell_r$. 

Equation (\ref{running G})  implies that the Hubble parameter \textit{today} is a scale-dependent quantity through $G(\ell)$, since one has

\begin{equation}\label{running H}
	H_0(\ell)^2 = \frac{8 \pi}{3} G(\ell)  \, \, \rho_0 ,
\end{equation}
where $\rho_0$ is the energy density of the universe today, that includes contributions from baryonic and dark matter, radiation, dark energy etc., while spatial curvature has been set to zero. For the sake of simplicity, we neglect the running of the vacuum energy $\rho_\Lambda$ in first instance, even though we have just mentioned that we expect it to be  a running quantity. For instance, in the  running vacuum model  this is  also required by the Bianchi identities in the Einstein's equations, see the discussion in \cite{pera 4,pera 5}. Nevertheless, it is easy to realize that a running vacuum energy would not spoil our arguments.

Equation (\ref{running H}) implies that measuring the Hubble parameter \textit{today}  at different scales $\ell$ would yield different results. Thus, if $\ell_{sn}$ the scale  probed by supernovae, while  $\ell_{cmb}$ is the scale probed by CMB, so that $H_0(\ell_{sn})=H_0^{sn}$ and  $H_0(\ell_{cmb})=H_0^{cmb}$, one gets a ratio

\begin{equation}\label{ratio H}
\left(H_0^{cmb}/H_0^{sn}\right)^2 = G\left(\ell_{cmb}\right)/G\left(\ell_{sn}\right) \, .
\end{equation}
Finally, as an estimation of $\ell_{sn}$ and $\ell_{cmb}$, one could take the distance traveled by light from supernovae and last scattering surface respectively, namely

\begin{equation}\label{ell sn cmb}	\begin{array}{ll}
		\ell_{sn} = a(z_{sn})\int_0^{z_{sn}} \, d\zeta/H(\zeta) 			\\
		\\
		\ell_{cmb} = a(z_{cmb})\int_0^{z_{cmb}} \,d\zeta/H(\zeta)
	\end{array}
\end{equation}
with $z_{sn}\sim 10^{-1}-1$ and $z_{cmb}\simeq 1100$.  Indeed, a scale-running  Newton's constant $G$ could address the Hubble tension.  We stress that the connection between the scale of renormalization with cosmological distances given in  $(\ref{ell sn cmb})$ is a reasonable  working hypothesis. However, although $(\ref{ell sn cmb}$) is not the only possible choice (e.g., alternatively, one could set $\ell_{sn}/\ell_{cmb} \simeq z_{cmb}$), the exact relation between $\mu$ and $\ell$ is not so relevant for the qualitative arguments given in this work. 

In support of such arguments, we mention that it has recently pointed our that a transition of $G$ at very late times/short distances would affect Cepheids and Type Ia Supernovae at distances less than $50 \, Mpc$, alleviating the Hubble tension \cite{melchiorri,ruchica}. In particular, in \cite{melchiorri}  it has been shown that a transition strength $\left[G(\ell_{sn})-G(\ell_{cmb})\right]/G(\ell_{sn}) \simeq 0.04$ is actually capable of solving the Hubble tension. However, even if the change in $G$ is considered as a local (short-scale) effect, in \cite{melchiorri,ruchica} this variation is introduced  by means of a time variation of the Newton's constant. On the contrary, in this letter we are proposing that the Hubble tension is actually an effect of the length-scale running of fundamental constant and cosmological parameters \textit{at the same time}.

In the following, we consider two different  theoretical frameworks, where the running of $G$ and $\Lambda$ comes from fundamental physics rather than from coarse graining of small cosmological perturbations.

In the first case, one  considers the effect of the zero-point energy of a  massive free scalar field on the cosmological constant, and derive the running of the vacuum energy density. This effect has been studied extensively in the past, and it is known under the name of running vacuum model (RVM hereafter) in the formulation presented in  \cite{casimir1,casimir2,casimir3,pera1,pera2,pera3,pera 4,pera 5,pera 6,pera 7,pera 8,pera H 1,pera H 2}.
However, we stress that in this paper we consider the vacuum energy density $\rho^0_\Lambda$ at present time as a running quantity that depends on the scale at which it is measured, instead of a time-varying quantity, as will be clarified below. Indeed, although the calculations are formally equal, this interpretation distinguish the proposal presented in this Letter from the RVM.
 
Though rigorous calculations have already been performed in \cite{casimir1,casimir2,casimir3,pera1,pera2,pera3,pera 4,pera 5,pera 6,pera 7,pera 8,pera H 1,pera H 2}, in order to  avoid technical complications that would make the discussion more cumbersome,  here we present a simplified heuristic derivation of the vacuum energy of a scalar field, which captures all the relevant physics involved.  Our starting point is the well known result that, in flat spacetime, the dimensionally regularized zero-point energy of a free massive scalar field is 

\begin{equation}\label{epsilon ZPE}
	\begin{array}{ll}
		\epsilon_{ZPE}= \mu^{4-n}\int \frac{dp^{n-1} }{(2 \pi \hbar)^3} \frac{\sqrt{p^2 c^2 + m^2 c^4}}{2} \, =\\
		\\
		= \frac{\beta}{2} \left(-\frac{2}{4-n}-\ln\left(\frac{4\pi \mu^2}{m^2 c^4}\right)+\gamma_E-\frac{3}{2}\right) \, ,
	\end{array}
\end{equation}
where $\gamma_E$ is the Euler's constant, $m$ is the mass of the scalar field, $\beta \equiv m^4 c^5/32 \pi^2 \hbar^3$, and $\mu$ is the energy scale at which $\epsilon_{ZPE}$ is evaluated, see  \cite{brown,casimir1,casimir2,casimir3}. This expression is divergent for $n\rightarrow 4$ and it must be renormalized. The contribution in equation (\ref{epsilon ZPE}) must be summed to the bare   cosmological constant. The latter can be splitted into its physical value $\rho^0_\Lambda$ plus a  counter-term $\delta\rho^0_\Lambda$  that  cancels the divergent part of $\epsilon_{ZPE}$, where the superscript $0$ means that all these quantities are evaluated at present cosmological times, that is at redshift $z=0$ .
Indeed, the renormalized vacuum energy density \textit{today} is

\begin{equation}\label{running rho lambda}
	\begin{array}{ll}
		\rho^0_\Lambda(\mu)\equiv  \rho^0_\Lambda + \delta \rho^0_\Lambda + \epsilon_{ZPE}=\\
		\\
		= 		\rho^0_\Lambda- \frac{\beta}{2}  \ln\left(\frac{\mu^2}{m^2 c^4}\right)= 		\rho^0_\Lambda(\mu_r)- \frac{\beta}{2}  \ln\left(\frac{\mu^2}{\mu_r^2}\right) \, .
	\end{array}
\end{equation} The last equality simply express the renormalized  vacuum energy density at the scale $\mu$ in terms of its  value at the generic renormalization scale $\mu_r$. 

As in the case of a running $G$ discussed above, the running vacuum energy density $\rho^0_\Lambda(\mu)$ enters the Hubble parameter by means of the Friedman equations, so that the value  $H_0$ of the Hubble parameter  today is also a running quantity that depends on the scale $\mu$. 

We assume for simplicity that the dark energy density is the only running parameter in the Friedman equations, indeed we neglect the running of $G$. We will motivate this assumption later on. Indeed,  the running  Hubble parameter today will be

\begin{equation}\label{H0 running}
	H_0^2\left(\mu\right) = \frac{8 \pi G}{3c^2}\left(\rho_\Lambda^0(\mu)+\rho^0_{matt}+\rho^0_{rad}+\ldots \right) \,,
\end{equation}
where $\rho^0_{matt}+\rho^0_{rad}+\ldots$ is the contribution to the energy density of the universe of matter, radiation etc., which is taken to be non-running to first approximation. By means of (\ref{running rho lambda}), one has

\begin{equation}\label{beta 1}
	\beta  \ln\left(\frac{\mu^2}{\mu_r^2}\right)= \frac{3 c^2}{4 \pi G} \left(H_0^2(\mu_r)-H_0^2(\mu)\right)\, .
\end{equation}
Setting $\mu_r$ as the "energy scale" corresponding to the Supernovae (this will be clarified right below), and $\mu$ as the "energy scale" associated to the CMB, one has $H_0(\mu)=H_0^{cmb}$
and \mbox{$H_0(\mu_r)=H_0^{sn}$}.

At that point, one can ask what is the physical meaning of the renormalization energy scale $\mu$ in the cosmological context. A reasonable working assumption is that $\mu$ must scale as the inverse of the size of length scale $\ell$ at which we observe the universe, that is $\mu \sim \hbar c/\ell$ and $\mu_r \sim \hbar c/\ell_r$. Indeed $\mu^2/\mu_r^2 \sim \ell_r^2/\ell^2\sim z_{cmb}^2 \sim 10^6$, as the universe  has expanded by factor of $z_{cmb}\sim 10^3$ since the last scattering surface of CMB. Putting numbers into (\ref{beta 1}), one obtains a value $m \sim  10^{-2} \, eV$, which is at the edge of the Axion mass range.  We note that the scalar field mass $m$  depends mildly on the exact value of the ratio $\ell/\ell_r$, as the dependence on $\mu/\mu_r$ in (\ref{beta 1}) is only logarithmic. We add that the possibility of a light neutrino in the RVM has also been discussed in \cite{casimir2}.

We stress that equation(\ref{running rho lambda}) have been obtained in a more rigorous fashion in  \cite{casimir1,casimir2,casimir3,pera1,pera2,pera3,pera 4,pera 5,pera 6,pera 7,pera 8,pera H 1,pera H 2}, where the renormalized vacuum energy has been expressed as

\begin{equation}\label{pera 1}
	\begin{array}{ll}
		\rho_\Lambda(H) = \rho^0_\Lambda + \delta \rho_\Lambda \\
		\\
		\delta\rho_\Lambda = \frac{3 }{8\pi}\, \nu_{eff}\,m_{Pl}^2
		\left(H^2-H_0^2\right)	\end{array} \, ,
\end{equation}
where $\nu_{eff} \sim \left(m/m_{Pl}\right)^2 \, \ln\left(m/m_{Pl}\right)^2$, and $m_{Pl}$ is the Planck mass. Equation (\ref{pera 1}) is the equivalent of equation (\ref{beta 1}), but in \cite{casimir1,casimir2,casimir3,pera1,pera2,pera3,pera 4,pera 5,pera 6,pera 7,pera 8,pera H 1,pera H 2} it has been given a completely different interpretation. In fact, $\rho_\Lambda(H)$ is considered as a time-varying quantity, and the effect of the vacuum renormalization of the scalar field has been studied by means of the inclusion of a time-varying vacuum energy density $\rho_\Lambda(H)$ in the Friedmann equations, which does not resolve the Hubble tension. On the contrary, in this paper we propose a completely different interpretation of the vacuum energy renormalization, namely that the vacuum energy density \text{today} $\rho^0_\Lambda$ is a scale-running quantity, which implies that measurements of the Hubble parameter \text{today} depend on the length scale at which the universe is probed. 

We mention that  the running of the Newton's constant  has been also evaluated in \cite{pera1,pera2,pera3}, giving  

\begin{equation}\label{pera 2}
	G(H) = G^0 /\left[1-\epsilon \ln\left(H^2/H_0^2\right)\right] \, ,
\end{equation}
where $\epsilon\sim m_i/m_{Pl}$, and $m_i$ receives contributions from all massive fields. Being logarithmic, such running is milder than that of $\Lambda$ given in equation (\ref{pera 1}), and this supports our assumption that the running of $G$ can be neglected in first approximation in this context.

We also mention that, to be rigorous, the heuristic arguments presented above should be generalized to QFT methods in curved spacetimes, as it  has been done in the case of the RVM in \cite{casimir1,casimir2,casimir3,pera1,pera2,pera3,pera 4,pera 5,pera 6,pera 7,pera 8,pera H 1,pera H 2}. Again, the formal calculations will be the same as in our case, while the interpretation of the running of $\Lambda$ and $G$ is different in this work.

For the second scenario, we analyze the case of the six derivative quantum gravity model studied in \cite{shapiro}, which assumes the following gravitational Lagrangian density

\begin{equation}\label{six derivative L}
	\begin{array}{ll}
		\mathcal{L} = \omega_\kappa R+ \omega_\Lambda+  \omega_C \, C_{\mu\nu\rho\sigma}  \Box C^{\mu\nu\rho\sigma}+
		\omega_R\,	R  \Box R  +\\
		\\+  \theta_R \, R^2 +
		\theta_C \, C^2+ \, \theta_{GB}  E_4
\end{array} \end{equation}
where $R$ is the scalar Ricci curvature, $E_4$ and $C^2$ are the Gauss-Bonnet and the Weyl terms respectively, which  are given by

\begin{equation}\label{weyl GB}
	\begin{array}{ll}
		C^2=R_{\mu\nu\rho\sigma} R^{\mu\nu\rho\sigma}-2 R_{\mu\nu}R^{\mu\nu}+R^2/3\\
		\\
		E_4= R_{\mu\nu\rho\sigma} R^{\mu\nu\rho\sigma}- 4R_{\mu\nu} R^{\mu\nu} + R^2 \,.
\end{array} \end{equation}

The model (\ref{six derivative L}) is a generalization of the minimal four derivative quadratic model studied long time ago in \cite{quadratic 1,quadratic 2,quadratic 3}, which corresponds to the choice $\omega_C= 0$ and $\omega_R=0$. In this sense, the case $\omega_C\neq 0$ and $\omega_R\neq0$ is non-minimal. Both the minimal and non-minimal models are renormalizable around a flat spacetime by means of the Barvinsky and Vilkovisky procedure \cite{barvisnki 1,barvisnki 2,barvisnki 3}. Furthermore,  higher than two derivative terms, that make the theory renormalizable, also entail the occurrence of ghosts. These must be removed from the physical spectrum using the so called Anselmi-Piva prescription \cite{anselmi 1,anselmi 2}. Alternatively, one can avoid the occurrence of ghosts  introducing nonlocal infinite-derivative terms \cite{nonlocal 1}.

Neglecting the problem of ghost states, we concentrate our attention on the renormalization scheme of (\ref{six derivative L}). It is well known that in the minimal model there is no running of $\omega_\Lambda$ and $\omega_\kappa$ \cite{donoghue,donoghue 2}, which are related to the cosmological constant $\Lambda$ and to the inverse of the Newton's constant $G$. However, this situation changes in the non-minimal model, where six derivative terms are introduced. In \cite{shapiro} the renormalization group equations for (\ref{six derivative L}) have been obtained explicitly at one loop for $\omega_C\neq 0$ and $\omega_R\neq0$, showing that the coefficients $\omega_C$ and $\omega_R$ of the six-derivative terms do not run, while all the other coefficients, including $\omega_\Lambda$ and $\omega_\kappa$, run with the renormalization scale. We skip some technicalities, e.g. the fact that the beta functions reported in \cite{shapiro} are correct at one loop level for $\omega_\Lambda$ and $\omega_\kappa$, but are exact at any loop for $\theta_R$, $\theta_C$, $\theta_{GB}$, and we focus on the running of $\Lambda$ and $G$. One has \cite{shapiro}

\begin{equation}\begin{array}{ll}\label{running shapiro}
		\frac{c^4}{16 \pi G(\mu)} \equiv \omega_\kappa(\mu) = \omega_\kappa(\mu_r) + a_1 \ln\left(\frac{\mu}{\mu_r}\right) + \\
		\\
		 a_2 \left[\ln\left(\frac{\mu}{\mu_r}\right)\right]^2, \\
		\\
		\Lambda(\mu)\equiv\omega_\Lambda(\mu) = \omega_\Lambda(\mu_r) + b_1 \ln\left(\frac{\mu}{\mu_r}\right) +   \\
		\\
 b_2 \left[\ln\left(\frac{\mu}{\mu_r}\right)\right]^2+		b_3 \left[\ln\left(\frac{\mu}{\mu_r}\right)\right]^3,
	\end{array}
\end{equation}
where $a_i$ and $b_i$ are non-running parameters, functions of the non running parameters $\omega_C$, $\omega_R$, and of the parameters $\omega_C$, $\omega_R$, $\theta_R$, $\theta_C$, $\theta_{GB}$ evaluated at the renormalization energy $\mu_r$. We refer the reader to \cite{shapiro} for the details, including the constraints on the initial values of the renormalization group equations that gives asymptotic freedom. What is relevant here is that equations (\ref{running shapiro}) give  the same running of $G$  as  in equation (\ref{running G}), with the identification $\ell \sim 1/\mu$. Indeed, the conjectured scale-dependence of the fundamental constants and cosmological parameters can be embedded in the non-minimal six derivatives model (\ref{six derivative L}).

Summarizing,  the main idea proposed in this work  is that the Hubble tension can be explained assuming that the fundamental constants and cosmological parameters in the standard $\Lambda$-CDM model are running quantities that depend on the scale at which they are measured. Consequently, also the value of the Hubble parameter today $H_0$ depends on the scale at which it is measured, indeed on the cosmological probes that are used for its estimation. 
Once again, we stress that the type of scale-running that we consider in this Letter is not reinterpreted as a time-dependence via a dependence of the renormalization scale on the Hubble rate or  the scale factor. On the contrary, we propose that fundamental constants and cosmological parameters take different values at the same time, depending on the scale at which they are measured.
We have argued that this proposal is supported by some observational evidence \cite{melchiorri,ruchica}, and explored different theoretical scenarios that can give rise to the right running of $G$ and $\Lambda$. The first one is the effective field theory of cosmological perturbations, where the running of fundamental constants and cosmological parameters is due to classical renormalization of the system, by means of the coarse graining of small scales inhomogeneities.  On the contrary, in the case of the RVM \cite{casimir1,casimir2,casimir3,pera1,pera2,pera3,pera 4,pera 5,pera 6,pera 7,pera 8,pera H 1,pera H 2} and the six-derivative model \cite{shapiro} the renormalization is genuinely quantum.

We conclude adding some final remarks. This paper is based on heuristic proof-of-concept arguments; however, this is enough for its scopes, i.e.,  to propose a new conceptual framework for addressing the Hubble tension. Of course, due to the preliminary nature of this paper, further work is needed to make this approach more robust. 	For instance, one has to quantify the effects of  the running of $G$ in the framework of the effective field theory of cosmological perturbations, and perform a detailed comparison with cosmological data, although preliminary arguments based on Ref.s \cite{ruchica,melchiorri} are encouraging. In this respect, we mention that the idea that local physical laws affect calibrators like Cepheids and Type Ia Supernovae is currently under active investigation, see e.g. \cite{Alestas,Perivolaropoulos}, and a route for testing this hypothesis  with next-generation facilities  is under consideration, see for instance the discussions in  \cite{Perivolaropoulos 2}. We just stress that the scenario of \textit{scale-varying} fundamental constants and cosmological parameters that has been proposed in this Letter can be relevant for this research line. For instance, in the context of the effective field theory of cosmological perturbations, one expects that a natural candidate for the step-varying transition length of $G$ is the size of non-linear short-wavelength perturbations. However, there might be more than one step-transitions of $G$ associated with different length-scales, or even a smooth quantum-induced scale-dependence as in equation (\ref{running G}). This would imply that, measuring $H_0$ with a different cosmological probe that tests the universe on a further cosmological scale, e. g. with primordial gravitational waves, one could get a  value of $H_0$ different from $H_0^{cmb}$ and $H_0^{sn}$.

\end{document}